\documentclass[11pt]{article}
\usepackage[letterpaper, margin=1in]{geometry}
\usepackage[margin=1cm]{caption}
\usepackage[utf8]{inputenc}
\usepackage{url, comment, amsmath, booktabs, graphicx, listings}
\usepackage{siunitx}
\usepackage{hyperref}
\usepackage{float}
\usepackage[dvipsnames]{xcolor}
\usepackage{authblk}

\usepackage[font={small,sf,bf}]{caption}

\newcommand\myurl[1]{\small{\url{#1}}}

\newcommand \numdatasets{23}
\newcommand \nummolecules{4.2~B}
\newcommand \datasize{60~TB}

\usepackage[backend=biber,style=nature,]{biblatex}
\newif\iffinal
\finaltrue

\iffinal
    \newcommand\ben[1]{}
    \newcommand\ian[1]{}
    \newcommand\xuefeng[1]{}
    \newcommand\kyle[1]{}
    \newcommand\status[1]{}
    \newcommand\note[1]{}
    \newcommand\todo[1]{}
\else
    \newcommand\ben[1]{{\color{MidnightBlue}[Ben: #1]}}
    \newcommand\ian[1]{{\color{red}[Ian: #1]}}
    \newcommand\xuefeng[1]{{\color{purple}[Xuefeng: #1]}}
    \newcommand\kyle[1]{{\color{green}[Kyle: #1]}}
    \newcommand\status[1]{{\color{blue}[Status \today{}: #1]}}
    \newcommand\note[1]{{\color{blue}[Note: #1]}}
    \newcommand\todo[1]{{\color{red}[TBD: #1]}}
\fi

\title{Targeting SARS-CoV-2 with AI- and HPC-enabled Lead Generation: A First Data Release}

\author[1,3]{Yadu Babuji}
\author[1,3]{Ben Blaiszik}
\author[4]{Tom Brettin}
\author[1,2,3]{Kyle Chard}
\author[1,5]{Ryan Chard}
\author[1,2]{Austin Clyde}
\author[1,2,3]{Ian Foster}
\author[1,2]{Zhi Hong}
\author[6,7]{Shantenu Jha}
\author[2]{Zhuozhao Li}
\author[2]{Xuefeng Liu}
\author[1,5]{Arvind Ramanathan}
\author[7]{Yi Ren}
\author[4]{Nicholaus Saint}
\author[1]{Marcus Schwarting}
\author[2,4]{Rick Stevens}
\author[7]{Hubertus van Dam}
\author[3]{Rick Wagner}
\author[ ]{other members of the AHLG-SARS-CoV-2 Collaboration}

\affil[1]{Data Science and Learning Division, Argonne National Laboratory}
\affil[2]{Department of Computer Science, University of Chicago}
\affil[3]{Globus, University of Chicago}
\affil[4]{Computing Environment and Life Sciences Directorate, Argonne National Laboratory}
\affil[5]{Consortium for Advanced Science and Engineering, University of Chicago}
\affil[6]{Electrical and Computer Engineering, Rutgers University}
\affil[7]{Computing Science Initiative, Brookhaven National Laboratory}

\begin{document}

\maketitle

\abstract{\noindent{}Researchers across the globe are seeking to rapidly repurpose existing drugs or discover new drugs to counter the %severe acute respiratory syndrome, also referred to as 
the novel coronavirus disease (COVID-19)
%COVID-19 is 
caused by severe acute respiratory syndrome coronavirus 2 (SARS-CoV-2).
One promising approach is to train machine learning (ML) and artificial intelligence (AI) tools to screen large numbers of small molecules.
As a contribution to that effort, we are aggregating numerous
small molecules from a variety of sources, using high-performance computing (HPC)
to computer diverse properties of those molecules, using the computed properties
to train ML/AI models, and then using the resulting models for screening.
%In particular, to spur the development of machine learning (ML) and artificial intelligence (AI) tools, it is critical to provide simple access to large collections of molecules specifically collected, curated, and enriched to speed this work.
%\ian{The value here is the computed properties, not the molecules per se, I think?}
%\kyle{Worth mentioning value for ML?}
In this first data release, we make available \numdatasets{} datasets collected from community sources representing over \nummolecules{} molecules enriched with pre-computed: 1) molecular fingerprints to aid similarity searches, 2) 2D images of molecules to enable exploration and application of image-based deep learning methods, and 3) 2D and 3D molecular descriptors to speed development of machine learning models. This data release encompasses structural information on the \nummolecules{} molecules and \datasize{} of pre-computed data.
Future releases will expand the data to include more detailed molecular simulations,
computed models, and other products.
}

\section{Introduction}
The Coronavirus Disease (COVID-19) pandemic, caused by transmissible infection of the severe acute respiratory syndrome coronavirus 2 (SARS-CoV-2) virus~\cite{zhou2020network, sheahan2020orally, Heiser2020.04.21.054387, Gordon2020}, has resulted in millions of diagnosed cases and over \num{353000} deaths worldwide~\cite{jhu-map}, straining healthcare systems, and disrupting key aspects of society and the wider economy. In order to save lives and reduce societal effects, it is important to rapidly find effective treatments through drug discovery and repurposing efforts.  

Here, we describe a public data release of \numdatasets{} molecular datasets collected from community sources or created internally, representing over \nummolecules{} molecules. In addition to collecting the datasets from heterogeneous locations and making them available through a unified interface, we have enriched the datasets with additional context that would be difficult for many researchers to compute without access to significant HPC resources. For example, these data now include the 2D and 3D molecular descriptors, computed molecular fingerprints, 2D images representing the molecule, and canonical simplified molecular-input line-entry system (SMILES)~\cite{weininger1989smiles} structural representations to speed development of machine learning models. 

This data release encompasses information on the \nummolecules{} molecules and \datasize{} of additional data. We intend to supplement this dataset in future releases with more datasets, further enrichments, tools to extract potential drugs from natural language text, and machine learning models to sift the best candidates for protein docking simulations from the billions of available molecules. In the following, we first describe the datasets collected, the methodology used to generate the enriched datasets, and then discuss future directions.

\section{Collected Datasets}
We have collected molecules from the datasets listed in Table \ref{tab:sources}, each of which has either been made available online by others or generated by our group. The collected datasets include some specifically collected for drug design (e.g., Enamine), known drug databases (e.g., Drugbank~\cite{DBK-article, DBK-web}, DrugCentral~\cite{DCL-article, DCL-web}, CureFFI~\cite{FFI-web}), antiviral collections (e.g., CAS COVID-19 Antiviral Candidate Compounds~\cite{CAS-web}, and the Lit COVID-19 dataset\cite{lit-db}), others that provide known decoys (DUDE database of useful decoys), and further counterexamples including molecules used in other domains (e.g., QM9~\cite{QM9-article, QM9-web}, Harvard Organic Photovoltaic Dataset~\cite{HOP-article, HOP-web}). By aggregating these diverse datasets, including the decoys and counterexamples, we aim to allow researchers the maximal freedom to create training sets for specific use cases. 
Future releases will include additional data relevant to SARS-CoV-2 research.

% Table of collected datasets and some descriptive statistics
\begin{table}[H]
    \centering
    \vspace{1ex}
\caption{The datasets included in the first data release, with for each a key, a brief description and references to the original location,  and the number of molecules.
%the fraction unique to only that dataset (\%U), 
Datasets labeled with \textsuperscript{$\dagger$} are provided as decoys or examples of molecules used in other domains.}\label{tab:sources}
    
    \begin{tabular}{l|l|r}
        \hline
        \textbf{Key} & \textbf{Name} & \textbf{\# Molecules}\\
        \hline
        BDB & \href{https://www.bindingdb.org/bind/index.jsp}{The Binding Database}~\cite{BDB-article, BDB-web} & \num{1813538}\\
        CAS & \href{https://www.cas.org/covid-19-antiviral-compounds-dataset}{CAS COVID-19 Antiviral Candidate Compounds}~\cite{CAS-web} & \num{49437}\\
        CHM & CheMBL db of bioactive mols with drug-like properties & \num{1940732} \\
        DBK & \href{https://www.drugbank.ca}{Drugbank}~\cite{DBK-article, DBK-web} & \num{9678}\\
        DCL & \href{http://drugcentral.org}{DrugCentral Online Drug Compendium}~\cite{DCL-article, DCL-web} & \num{3981}\\
        DUD & \href{http://dude.docking.org}{\textsuperscript{$\dagger$}DUDE database of useful decoys}~\cite{DUD-article, DUD-web} & \num{99782}\\
        E15 & \href{https://enamine.net/library-synthesis/real-compounds/real-compound-libraries}{Diverse REAL drug-like subset of ENA} & \num{15547091}\\
        EDB & \href{https://enamine.net/hit-finding/diversity-libraries/hit-locator-library-300}{DrugBank plus Enamine Hit Locator Library 2018}~\cite{EDB-web} & \num{310782}\\
        EMO & \href{https://www.emolecules.com/info/products-data-downloads.html}{eMolecules}~\cite{EMO-web} & \num{25946988}\\
        ENA & \href{https://enamine.net/library-synthesis/real-compounds/real-database}{Enamine REAL Database}~\cite{ENA-article, ENA-web} & \num{1211723723}\\
        FFI & \href{https://www.cureffi.org/2013/10/04/list-of-fda-approved-drugs-and-cns-drugs-with-smiles/}{CureFFI FDA-approved drugs and CNS drugs}~\cite{FFI-web} & \num{1497}\\
        G13 & \href{http://gdb.unibe.ch/downloads/}{GDB-13 small organic molecules up to 13 atoms}~\cite{G13-article, G13-web} & \num{977468301}\\
        G17 & \href{http://gdb.unibe.ch/downloads/}{GDB-17-Set up to 17 atom extension of GDB-13}~\cite{G17-article, G17-web} & \num{50000000}\\
        HOP & \href{https://www.nature.com/articles/sdata201686}{\textsuperscript{$\dagger$}Harvard Organic Photovoltaic Dataset}~\cite{HOP-article, HOP-web} & 350\\
        LIT & COVID-relevant small mols extracted from literature~\cite{lit-db}& 803\\
        MOS & \href{https://github.com/molecularsets/moses}{Molecular Sets (MOSES)}~\cite{MOS-article, MOS-web} & \num{1936962}\\
        MCU & MCULE compound database & \num{45472755}\\
        PCH & \href{https://www.ncbi.nlm.nih.gov/guide/data-software/}{PubChem}~\cite{PCH-article, PCH-web} & \num{97545266}\\
        QM9 & \href{http://quantum-machine.org/datasets/}{QM9 subset of GDB-17}~\cite{QM9-article, QM9-web} & \num{133885}\\
        REP & \href{https://clue.io/data/REP#REP}{Repurposing-related drug/tool compounds}~\cite{REP-article, REP-web} & \num{10141}\\
        SAV & \href{https://cactus.nci.nih.gov/download/savi_download/}{Synthetically Accessible Virtual Inventory (SAVI)}~\cite{SAV-article, SAV-web} & \num{265047097}\\
        SUR & \href{https://surechembl.org/}{SureChEMBL dataset of molecules from patents}~\cite{SUR-article, SUR-web} & \num{17915384}\\
        ZIN & \href{http://zinc15.docking.org}{ZINC15}~\cite{ZIN-article, ZIN-web} & \num{1225804829}\\
        \hline
       \multicolumn{2}{l|}{\textbf{Total}} & \textbf{\num{4206934042}}\\
        \hline
    \end{tabular}
        
\end{table}

\section{Methodology and Data Processing Pipeline}

The data processing pipeline is used to compute different types of features and representations of billions of small molecules. The pipeline is first used to convert the SMILES representation for each molecule to a canonical SMILES to allow for de-duplication and consistency across data sources. Next, for each molecule, three different types of features are computed: 1) molecular fingerprints that encode the structure of molecules; 2) 2D and 3D molecular descriptors; and 3) 2D images of the molecular structure. These features are being used as input to various machine learning and deep learning models that will be used to predict important characteristics of candidate molecules including docking scores, toxicity, and more.

% Graphic showing the pipeline. The generator for this image is in the /img/generators folder
\begin{figure}[H]
    \begin{center}
        \includegraphics[width=0.88\columnwidth,trim=6mm 40mm 6mm 68mm,clip]{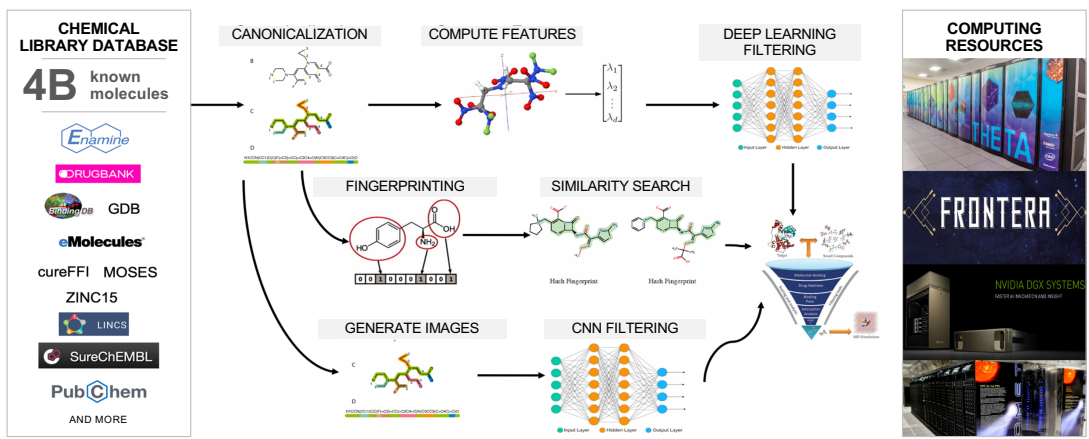}
    \end{center}
    
    \vspace{-2ex}
    
    \caption{The computational pipeline that is used to enrich the data collected from included datasets. After collection, each molecule in each dataset has canonical SMILES, 2D and 3D molecular features, fingerprints, and images computed. These enrichments simplify molecule disambiguation, ML-guided compound screening,  similarity searching, and neural network training respectively.}
    \label{fig:pipeline}
\end{figure}

% Table describing key terms used in the methodology section
\begin{table}[htb!]
\centering
    %\vspace{1ex}
        \caption{Definitions for terms used in the methodology section to describe key aspects of the collected datasets and computed properties.}
    \label{tab:terms}
    
    \begin{tabular}{l|p{11.4cm}}
        \hline
        \textbf{Term} & \textbf{Description} \\
        \hline
        SOURCE-KEY & Identifies the source dataset: see the three-letter \textbf{``Keys''} in Table \ref{tab:sources} \\
        IDENTIFIER & A per-molecule identifier either obtained from the source dataset or, if none such is available, defined internally \\
        SMILES & A canonical SMILES for a molecule, as produced by Open Babel \\ \hline
    \end{tabular}
    \centering
\end{table}

\subsection{Canonical Molecule Structures}

We use Open Babel v3.0~\cite{o2011open} to convert the simplified molecular-input line-entry system (SMILES) specifications of chemical species obtained from various sources into a consistent canonical smiles representation. We organize the resulting molecule specifications in one directory per source dataset, each containing one CSV file with columns $[$SOURCE-KEY, IDENTIFIER, SMILES$]$, where SOURCE-KEY identifies the source dataset; IDENTIFIER is an identifier either obtained from the source dataset or, if none such is available, defined internally; and SMILES is a canonical SMILES as produced by Open Babel.  Identifiers are unique within a dataset, but may not be unique across datasets. Thus, the combination of (SOURCE-KEY, IDENTIFIER) is needed to identify molecules uniquely. We obtain the canonical SMILES by using the following Open Babel command:

 \begin{center}
\texttt{obabel \{input\_filename\} -O \{output\_filename\} -ocan -e}
 \end{center}

\subsection{Molecular Fingerprints}

\begin{figure}[H]
    \begin{center}
        \fbox{\includegraphics[width=0.87\textwidth,trim=0 4mm 8mm 0,clip]{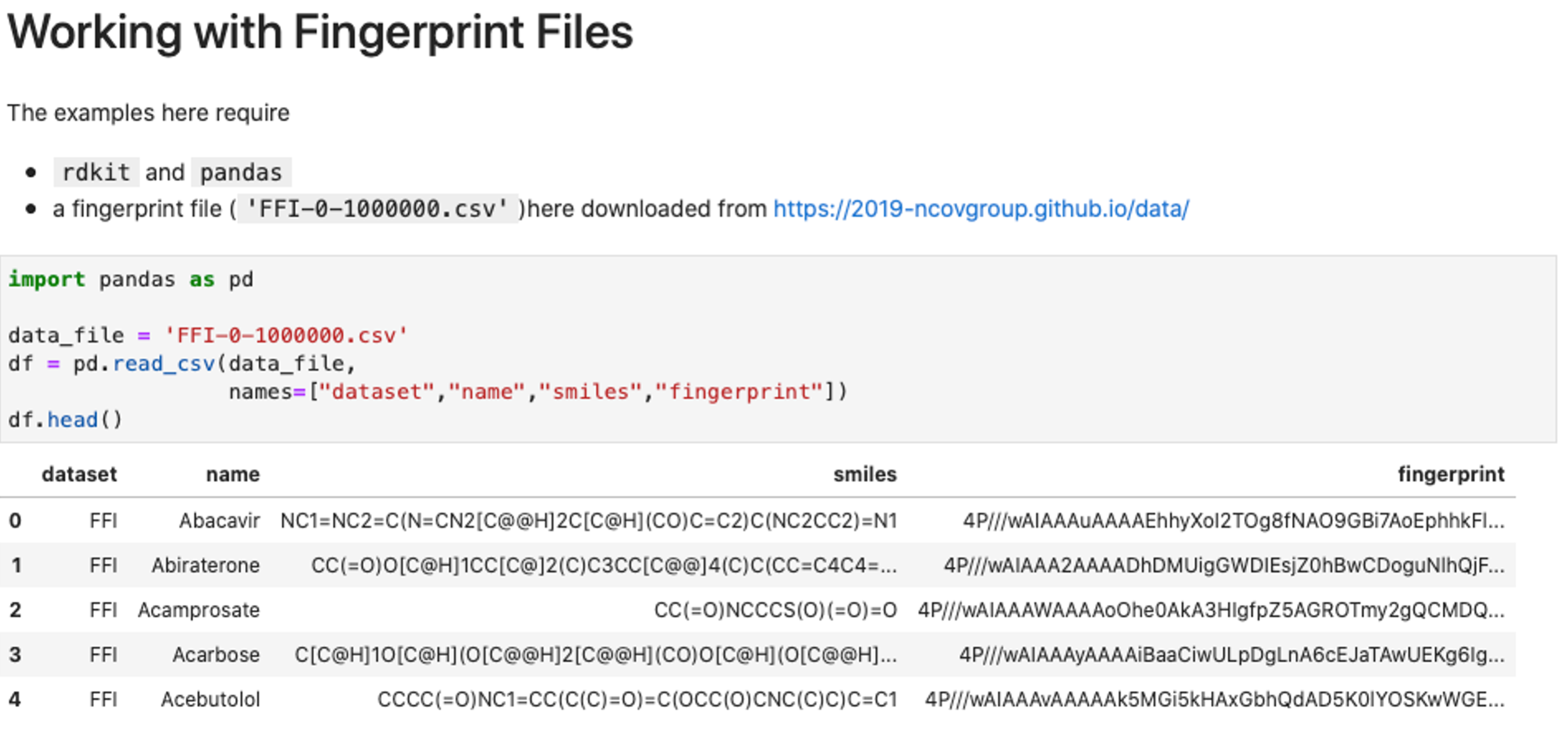}}
    \end{center}
    
        \vspace{-2ex}
        
    \caption{A simple Python code example showing how to load data from a fingerprint file. (This and other examples are accessible on GitHub~\cite{covid-analyses-repo}.)}
    \label{fig:fingerprints}
\end{figure}

We use RDKit~\cite{landrum2013rdkit} (version 2019.09.3) to compute a 2048-bit fingerprint for each molecule. We organize these fingerprints in CSV files with each row with columns $[$SOURCE-KEY, IDENTIFIER, SMILES, FINGERPRINT$]$, where SOURCE-KEY, IDENTIFIER, and SMILES are as defined in Table \ref{tab:terms}, and FINGERPRINT is a Base64-encoded representation of the fingerprint. In Figure~\ref{fig:fingerprints}, we show an example of how to load the fingerprint data from a batch file within individual dataset using Python 3. Further examples of how to use fingerprints are available in the accompanying GitHub repository~\cite{covid-analyses-repo}.

\subsection{Molecular Descriptors}

\begin{figure}[H]
    \begin{center}
        \fbox{\includegraphics[width=0.87\textwidth,trim=0 4mm 23mm 4mm,clip]{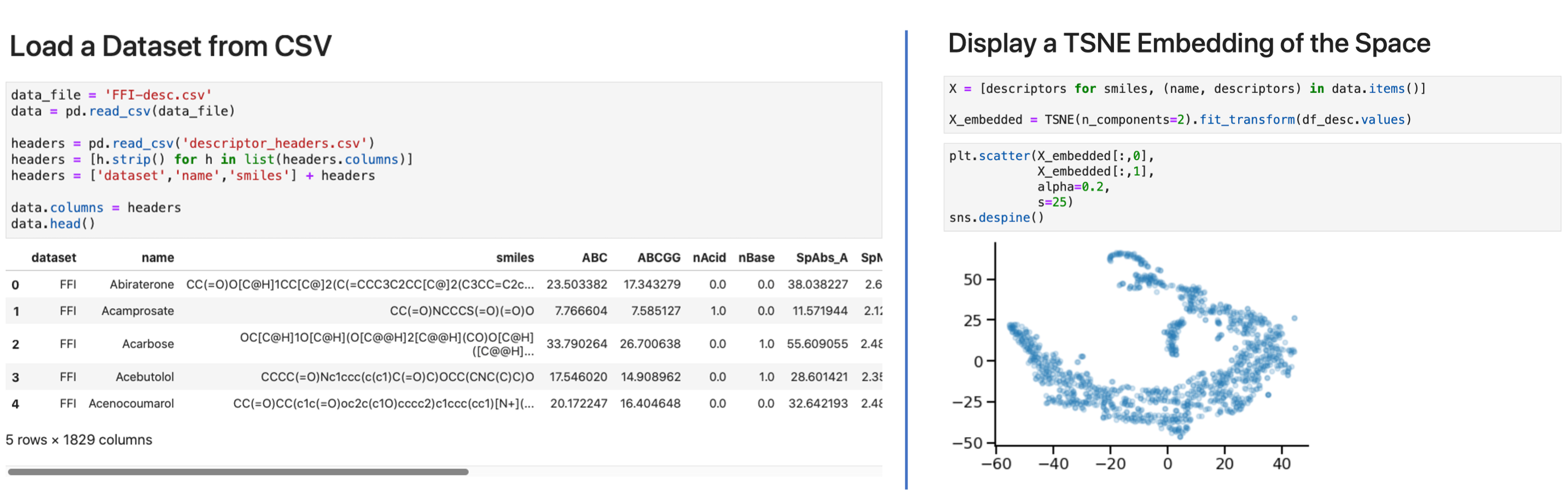}}
    \end{center}
    
    \vspace{-2ex}
    
    \caption{Molecular descriptor examples: (left) load descriptor data and (right) create a simple TSNE projection of the FFI dataset.}
    \label{fig:descriptors}
\end{figure}

We generate molecular descriptors using Mordred~\cite{moriwaki2018mordred} (version 1.2.0). The collected descriptors ($\sim$1800 for each molecule) include descriptors for both 2D and 3D molecular features. We organize these descriptors in one directory per source dataset, each containing one or more CSV files. 
Each row in the CSV file has columns $[$SOURCE-KEY, IDENTIFIER, SMILES, DESCRIPTOR$_1$ ... DESCRIPTOR$_N]$.
In Figure~\ref{fig:descriptors}, we show how to load the data for an individual dataset (e.g., FFI) using Python 3 and explore its shape (Figure~\ref{fig:descriptors}-left), and  create a TSNE embedding~\cite{maaten2008visualizing} to explore the molecular descriptor space (Figure~\ref{fig:descriptors}-right). 

\subsection{Molecular Images}
\begin{figure}[H]
    \begin{center}
        \fbox{\includegraphics[width=0.87\textwidth,trim=0 2mm 4mm 0,clip]{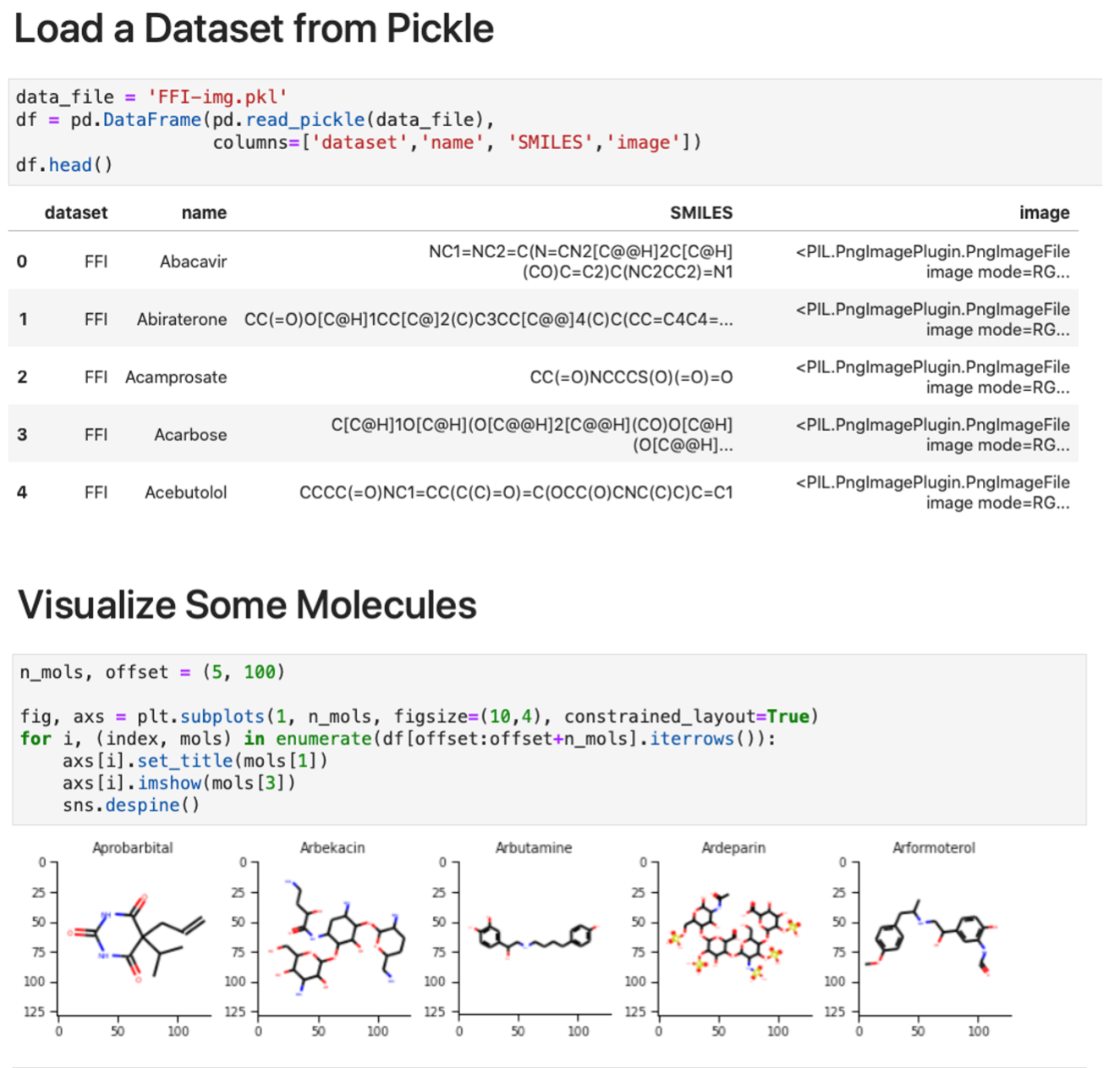}}
    \end{center}
    
    \vspace{-2ex}
    
    \caption{Molecular image examples. The examples show how to (top) load the data and (bottom) display a subset of the images using matplotlib.}
    \label{fig:images}
\end{figure}

Images for each molecule were generated using a custom script~\cite{covid-analyses-repo} to read the canonical SMILES structure with RDKit, kekulize the structure, handle conformers, draw the molecule with rdkit.Chem.Draw, and save the file as a PNG-format image with size 128$\times$128 pixels. For each dataset, individual pickle files are saved containing batches of \num{10000} images for ease of use, with entries in the format (SOURCE, IDENTIFIER, SMILES, image in PIL format). In Figure~\ref{fig:images}, we show an example of loading and display image data from a batch of files from the FFI dataset. 

\section{Data Access}
\begin{figure}[H]
    \begin{center}
        \fbox{\includegraphics[width=0.87\textwidth]{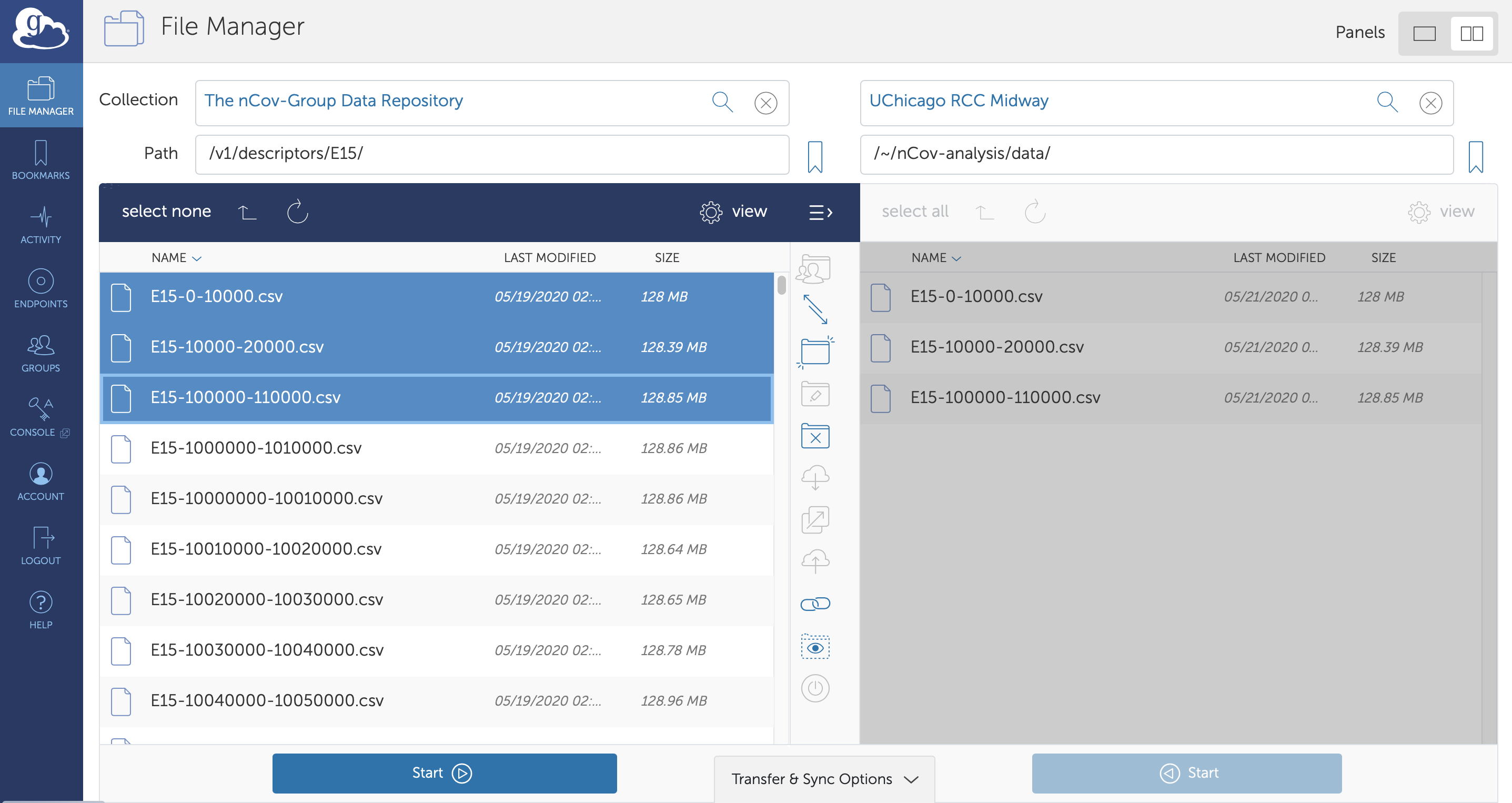}}
    \end{center}
    
    \vspace{-2ex}
    
    \caption{Data access with Globus. All data are stored on Globus endpoints, allowing users to access, move, and share the data through a web interface (pictured above), a REST API, or with a Python client. The user here has just transferred the first three files of descriptors associated with the E15 dataset to an endpoint at UChicago.
    \label{fig:globus}}
\end{figure}

Providing access to such a large quantity of heterogeneous data (currently~\datasize{}) is challenging. We use Globus~\cite{chard2016globus} to handle authentication and authorization, and to enable high-speed, reliable access to the data stored on the Petrel file server at the Argonne Leadership Computing Facility's (ALCF) Joint Laboratory for System Evaluation (JLSE). Access to this data is available to anyone following authentication via institutional credentials, an ORCID profile, a Google account, or many
other common identities. Users can access the data through a web user interface shown in Fig.~\ref{fig:globus}, facilitating easy browsing, direct download via HTTPS of smaller files, and high-speed, reliable transfer of larger data files to their laptop or a computing cluster via Globus Connect Personal or an instance of Globus Connect Server. There are more than \num{20000} active Globus endpoints distributed around the world.  Users may also access the data with a full-featured Python SDK. More details on Globus can be found at \url{https://www.globus.org}.

\section{Conclusion and Future Directions}
We have released to the community an open resource of molecular structures (as canonical SMILES), descriptors, 2D images, and fingerprints. We hope these data will contribute to  the discovery of small molecules to combat the SARS-CoV-2 virus.

We expect forthcoming data releases to extend to molecular conformers; incorporate the results of natural language processing extractions of drugs from COVID-related literature; provide the results of molecular docking simulations against SARS-CoV-2 viral and host proteins; and include the trained machine learning models that the team is building to identify top candidates for running various, more expensive calculations. % predict toxicity, and more.

\section{Data and Code Availability}
All data and code links can be found at \url{http://2019-ncovgroup.github.io/data/}. Subsequent updates will be made available through the same web page, and further release papers will be issued as necessary. The code for the examples used in this paper can be found at \url{https://github.com/globus-labs/covid-analyses}.

\section{Acknowledgements}

The data generated have been prepared as part of the nCov-Group Collaboration, a group of over 200 researchers working to use computational techniques to address various challenges associated with COVID-19. We would like to thank all the researchers who helped to assemble the original datasets, and who provided permission for redistribution.

This research was supported by the DOE Office of Science through the National Virtual Biotechnology Laboratory, a consortium of DOE national laboratories focused on response to COVID-19, with funding provided by the Coronavirus CARES Act. This research used resources of the Argonne Leadership Computing Facility, a DOE Office of Science User Facility supported under Contract DE-AC02-06CH11357. Additional data storage and computational support for this research project has been generously supported by the following resources: Petrel Data Service at the Argonne Leadership Computing Facility (ALCF), Frontera at the Texas Advanced Computing Center (TACC), Comet at the San Diego Supercomputing Center (SDSC)

The work leveraged data and computing infrastructure produced in other projects, including: ExaLearn and the Exascale Computing Project~\cite{alexander2020exascale} (DOE Contract DE-AC02- 06CH11357); Parsl: parallel scripting library~\cite{babuji2019parsl} (NSF 1550588); funcX: distributed function as a service platform~\cite{chard2019serverless} (NSF 2004894); Globus: data services for science (authentication, transfer, users, and groups (see \url{globus.org} for funding); CHiMaD: Materials Data Facility~\cite{blaiszik2019data, blaiszik2016materials} and Polymer Property Predictor Database~\cite{tchoua2019creating} (NIST 70NANB19H005 and NIST 70NANB14H012)

\section{Disclaimer}

For All Information. Unless otherwise indicated, this information has been authored by an employee or employees of the UChicago Argonne, LLC, operator of the Argonne National laboratory with the U.S.\ Department of Energy. The U.S.\ Government has rights to use, reproduce, and distribute this information. The public may copy and use this information without charge, provided that this Notice and any statement of authorship are reproduced on all copies.

While every effort has been made to produce valid data, by using this data, User acknowledges that neither the Government nor UChicago Argonne, LLC, makes any warranty, express or implied, of either the accuracy or completeness of this information or assumes any liability or responsibility for the use of this information. Additionally, this information is provided solely for research purposes and is not provided for purposes of offering medical advice. Accordingly, the U.S.\ Government and UChicago Argonne, LLC, are not to be liable to any user for any loss or damage, whether in contract, tort (including negligence), breach of statutory duty, or otherwise, even if foreseeable, arising under or in connection with use of or reliance on the content displayed on this site.

\printbibliography

\end{document}